\documentclass[aps,twocolumn,amsmath,amssymbs,superscriptaddress,longbibliography
 amsmath,amssymb,
 aps,
]{revtex4-2}

\usepackage{graphicx}
\usepackage{dcolumn}
\usepackage{bm}
\usepackage{color}

\newcommand{\braket}[1]{\langle #1 \rangle}
\definecolor{colP1}{rgb}{0.0, 0.3, 0.8}
\definecolor{colP2}{rgb}{0.7, 0.0, 0.7}
\definecolor{colP3}{rgb}{0.85, 0.5, 0.0}
\definecolor{colP4}{rgb}{0.75, 0.0, 0.0}
\definecolor{colP5}{rgb}{0.0, 0.5, 0.0}
\definecolor{colP6}{rgb}{0.4, 0.4, 0.4}

\usepackage{hyperref}

\begin{document}

\preprint{APS/123-QED}

\title{Continuous-variable model for arbitrary image propagation\\via Dirac-comb expansion in Four-Wave Mixing}

\author{Fabián Ramírez-Pacheco}
\email{fabian.ramirezpacheco@gmail.com}
\affiliation{%
 Departamento de Física, Facultad de Ciencias Físicas y Matemáticas, Universidad de Chile, Santiago,
Chile
}%
\affiliation{
 Millennium Institute for Research in Optics - MIRO, Santiago, Chile
}%
\author{Andrea Basilio-Zárate}%
\email{andrea.basilio@ug.uchile.cl}
\affiliation{%
 Departamento de Física, Facultad de Ciencias Físicas y Matemáticas, Universidad de Chile, Santiago,
Chile
}%
\affiliation{
 Millennium Institute for Research in Optics - MIRO, Santiago, Chile
}

\author{Hans Marin Florez}
\affiliation{
 Centro de Ci\^encias Naturais e Humanas, Universidade Federal do ABC--UFABC, Santo Andr\'e 09210-580, Brazil
}%

\author{Pablo Solano}
\affiliation{
 Departamento de Física, Facultad de Ciencias Físicas y Matemáticas, Universidad de Concepción, Concepción, Chile
}%

\author{Carla Hermann-Avigliano}
\affiliation{%
 Departamento de Física, Facultad de Ciencias Físicas y Matemáticas, Universidad de Chile, Santiago,
Chile
}%
\affiliation{
 Millennium Institute for Research in Optics - MIRO, Santiago, Chile
}%

\date{\today}

\begin{abstract}
We develop a macroscopic continuous-variable model of stimulated four-wave mixing (4WM) yielding closed-form expressions for the mean intensity, variance, and covariance of bright probe and conjugate beams, valid for arbitrary transverse profiles of the pump and seed to all orders in the nonlinear interaction strength. The model uses a Dirac-comb expansion of the pump field, making the photon statistics computable pixel-by-pixel from one set of expressions. In the plane-wave limit, we recover the phase-insensitive amplifier results. Beyond this limit, the formalism explicitly demonstrates the spatial routing of information observed in earlier experiments: the seed angular spectrum is transferred to the twin-beam intensities, whereas the angular spectrum of the squared pump field is transferred to the spatial cross-correlation between the probe and conjugate. We characterize the parameter window of high-fidelity image transfer in the covariance, which is controlled by the seed-to-pump waist ratio and the interaction strength. Our model provides, in one analytical framework, a theoretical account of experiments encoding information in the bright intensities or non-locally in the spatial cross-correlations.
\end{abstract}

\maketitle

\section{Introduction}

The transverse spatial degrees of freedom of light have emerged as a central resource for high-dimensional quantum information~\cite{Nielsen2012,Horodecki2009,Otte2020}. Encoding information in spatial modes, such as orbital angular momentum, transverse pixels, or arbitrarily structured profiles, provides access to a continuous, high-dimensional Hilbert space, enabling parallel encoding, larger channel capacities, and richer correlation structures than polarization or single-mode platforms~\cite{Vaziri2002,Sit2018,Braunstein2005,Defienne2024}. Quantum-correlated bright twin beams generated by parametric processes are natural carriers of such spatially structured states because the intensity profile and nonlocal correlation between the beams can be tailored at the source~\cite{Giovannetti2011,Kolobov2007}.


Four-wave mixing (4WM) in alkali-atom vapors has become one of the most widely used platforms for generating bright twin beams owing to the strong resonant third-order nonlinearity of the medium~\cite{McCormick2006,Boyer2008}. In 4WM, a strong classical pump beam drives a $\chi^{(3)}$ interaction, in which two pump photons are converted into a probe--conjugate photon pair under energy and momentum conservation~\cite{Glorieux2010,Turnbull2013}. When the process is seeded by injecting light into the probe mode, it operates as a parametric amplifier, producing macroscopic mutually correlated beams that exhibit strong intensity-difference squeezing~\cite{McCormick2006,Boyer2008,Jasperse2011,Fang2015} and position--momentum EPR entanglement~\cite{Kumar2021}. Bright twin beams have also been generated in cold atomic ensembles, where the observed quantum correlations are consistent with non-perturbative theoretical predictions~\cite{Lambrecht1996,Arajo2022}, indicating that effects beyond the lowest-order interaction are accessible in current experiments.


Several experimental configurations have explored the spatial control of the pump and seed to encode information in the intensity profiles of the twin beams or their nonlocal spatial correlations. Seed shaping in 4WM faithfully transfers an arbitrary input image to the bright probe and conjugate intensities~\cite{Boyer2008}, whereas pump shaping has been used to imprint structured profiles directly in the spatial cross-correlation between the probe and conjugate, an encoding that is inaccessible by either beam individually~\cite{Nirala2023}. Related demonstrations include the transfer of orbital angular momentum from the pump to the generated modes~\cite{Walker2012,Offer2018,Yu2023} and slit-shaped pump amplitude profiles~\cite{Swaim2018}. 
These experiments established the bright-beam regime as a flexible platform for both local and nonlocal spatial encoding; however, their theoretical description has so far rested either on perturbative single-pair models or case-specific calculations.

The spectral and temporal properties of the 4WM-generated twin beams are well understood~\cite{Barbosa2020,He2020,Celis2025,Celis2025-arxiv}, but a unified description of their transverse spatial statistics in the bright-beam regime remains open. Two complementary lines of theoretical work have developed in parallel without convergence. On the one hand, biphoton-state treatments of structured-pump parametric processes, both in spontaneous parametric down-conversion (SPDC)~\cite{Monken1998,Walborn2010,Karan2020,Boucher2021} and, more recently, in 4WM~\cite{Thachil2025,daMotta2025}, accommodate arbitrary pump and seed profiles but rely on a first-order perturbative expansion of the interaction Hamiltonian. They characterize the field through the two-photon wavefunction or the coincidence rate, which are observables natural to the photon-pair regime rather than to bright continuous-mode beams. On the other hand, all-order Bogoliubov treatments of 4WM in the plane-wave limit yield closed-form expressions for the mean intensity, variance, covariance, and intensity-difference squeezing, but only for a single transverse mode pair~\cite{Jasperse2011,Fang2015}, discarding the high-dimensional Hilbert space of the transverse degrees of freedom. Consequently, no current formalism simultaneously (i) admits arbitrary transverse profiles for both the pump and seed, (ii) is valid for all orders in the interaction strength, and (iii) yields closed-form expressions for the bright-beam intensities, variances, and covariances from a single derivation.

In this work, we develop a macroscopic continuous-variable model that closes the gap described above. Starting from the 4WM interaction Hamiltonian, we derive Bogoliubov-type transformations of the probe and conjugate annihilation operators for an arbitrary classical pump profile and use them to obtain closed-form expressions for the mean intensity, variance of each beam, cross-covariance, and intensity-difference squeezing, all valid to all orders in the interaction strength. Expanding the pump field on a Dirac-comb basis renders the photon statistics computable pixel-by-pixel, making the formalism directly applicable to arbitrary input images. In the plane-wave limit, the model reproduces the phase-insensitive amplifier results, including the intensity-difference squeezing factor $1/(2G-1)$ with gain $G=\cosh^2 s$. Beyond this limit, the same expressions explicitly describe how the angular spectrum of the seed is mapped onto the intensities of the twin beams and how the angular spectrum of the squared pump field is mapped onto their spatial cross-correlation. This provides an all-order macroscopic theoretical account of the image-transfer experiments of Refs.~\cite{Boyer2008} and~\cite{Nirala2023}. We further characterize the parameter window of high-fidelity image transfer in the covariance, controlled by the seed-to-pump waist ratio $\delta$ and the interaction strength $s$, and identify $\delta\ll 1$ and $s\lesssim 1$ as the regime in which an arbitrary image imprinted on the pump is faithfully recovered in the cross-correlation.

The remainder of this paper is organized as follows. In Sec.~\ref{sec model}, we derive the general macroscopic framework for the photon statistics in 4WM and verify its consistency in the plane-wave limit. In Sec.~\ref{sec Dirac Comb}, we introduce the Dirac-comb expansion of the pump, apply it to image transfer in both the intensities and the spatial cross-correlation, and analyze the fidelity and parameter dependence of the covariance imaging. Finally, we summarize our conclusions in Sec.~\ref{sec conclusions}.

\section{\label{sec model}Theoretical Model}
In this section, we describe the 4WM in terms of an electric field interacting with a nonlinear medium. We derive the output relations for the bosonic operators of the probe and conjugate fields and present their statistical properties, such as the mean value, variance, and covariance of the intensities. Finally, we evaluate the plane-wave pump limit and verify the consistency of the model with the phenomenological description.

\subsection{4WM Hamiltonian and operator transformations}
The Hamiltonian of the electromagnetic field interacting with an arbitrary $\boldsymbol{\chi}^{(n)}$ medium is given by \cite{Drummond2014}
\begin{equation}
    H = \int \left[ \frac{\epsilon_0|\textbf{E}|^2}{2} + \frac{|\textbf{B}|^2}{2\mu_0}  + \sum_{n\geq 1}\frac{\epsilon_0 n}{n+1}\textbf{E}\cdot\boldsymbol{\chi}^{(n)}:\textbf{E}^{\otimes n}\right]\text{d}^3r.
    \label{ham int E}
\end{equation}
For our particular case, we consider $n=3$ and the product of the fields that contributed to the 4WM, $\hat{\textbf{E}}^{(+)}_p(\textbf{r},t)\hat{\textbf{E}}^{(+)}_p(\textbf{r},t)\hat{\textbf{E}}^{(-)}_{pr}(\textbf{r},t)\hat{\textbf{E}}^{(-)}_c(\textbf{r},t)$, and its conjugate process. The pump is considered a classical strong beam, which can be expressed as 
\begin{equation}
	\hat{\textbf{E}}^{(+)}_p(\textbf{r},t) \to \textbf{E}^{(+)}_p(\textbf{r},t) = E_p(\boldsymbol{\rho})e^{i(k_p^z z - \omega_p t)},
\end{equation}
where we use cylindrical coordinates $\textbf{r} = (\boldsymbol{\rho},z)$, with the pump field propagating along $z$ with an optical frequency $\omega_p$. The probe and conjugate fields can be considered quantum fields with optical frequencies $\omega_{pr}$ and $\omega_{c}$ which are described as
\begin{align}
	\hat{\textbf{E}}_{pr}^{(-)}(\textbf{r},t) &= \frac{-i}{2\pi}\sqrt{\dfrac{\hbar\omega_{pr}}{2 \epsilon_0 }}\int \hat{a}^\dagger_{\textbf{q}_{pr}}e^{-i(\textbf{k}_{pr}\cdot\textbf{r} - \omega_{pr} t)}\text{d}\textbf{q}_{pr},\\
    \hat{\textbf{E}}_{c}^{(-)}(\textbf{r},t) &= \frac{-i}{2\pi}\sqrt{\dfrac{\hbar\omega_c}{2 \epsilon_0 }}\int \hat{b}^\dagger_{\textbf{q}_{c} }e^{-i(\textbf{k}_c\cdot\textbf{r} - \omega_{c} t)}\text{d}\textbf{q}_{c},
\end{align}
with the notation $\textbf{k} = (\textbf{q}, k^z)$, where $\textbf{q}$ represents the transverse wave vector. Assuming energy conservation, the optical frequencies satisfy $2\omega_p = \omega_{pr} + \omega_c$, therefore the Hamiltonian takes the following form \cite{Thachil2025}
\begin{equation}
	\hat{\mathcal{H}} = i\hbar\Gamma \int \mathrm{F}(\textbf{q}_{pr},\textbf{q}_c)\hat{a}^\dagger_{\textbf{q}_{pr}}\hat{b}^\dagger_{\textbf{q}_c}\text{d}\textbf{q}_{pr}\text{d}\textbf{q}_c + \text{h.c.},
    \label{ham}
\end{equation}
with the two-photon amplitude (TPA) given by
\begin{equation}
    \mathrm{F}(\textbf{q}_{pr},\textbf{q}_c) = \mathcal{F}\left[E_p^2\right](\textbf{q}_{pr}+ \textbf{q}_c) \text{ sinc}\left( \frac{\Delta k^z L}{2}\right),\label{TPA}
\end{equation}\\
where $\Gamma = 3\epsilon_0\chi^{(3)}\sqrt{\omega_{pr}\omega_c}L/16\pi$ is the \textit{coupling parameter}, $\mathcal{F}\left[\cdot\right]$ is the Fourier transform operation, $\Delta k^z = 2k_p^z - k_{pr}^z - k_c^z$ is the longitudinal mismatch, and $\hat{a}^\dagger_{\textbf{q}_{pr}}$ ($\hat{b}^\dagger_{\textbf{q}_c}$) is the creation operator of a photon in the probe (conjugate) mode with transverse wave vector $\textbf{q}_{pr}$ ($\textbf{q}_{c}$). The numerical prefactor in $\Gamma$ depends on the convention used to normalize the field operators. In the following, $\Gamma$ enters only through the dimensionless interaction strength $s$ defined below. The probe and conjugate operators satisfy the commutation relations
\begin{equation}
    [\hat{a}_{\textbf{q}_1}, \hat{a}_{\textbf{q}_2}^\dagger] = [\hat{b}_{\textbf{q}_1}, \hat{b}_{\textbf{q}_2}^\dagger] = \delta(\textbf{q}_1 - \textbf{q}_2).
\end{equation}
Under phase-matching conditions, the TPA corresponds to the convolution of the two angular spectra of the pump. Based on the interaction Hamiltonian in Eq.~\eqref{ham} the evolution operator is 
\begin{equation}
    \hat{U} = \exp\left[\Gamma t \int \mathrm{F}(\textbf{q}_{pr},\textbf{q}_c)\hat{a}^\dagger_{\textbf{q}_{pr}}\hat{b}^\dagger_{\textbf{q}_c}\text{d}\textbf{q}_{pr}\text{d}\textbf{q}_c - \text{h.c.}\right],
\end{equation}
where $t$ is the interaction time. By applying the unitary transformation to the quantum fields with transverse direction $\textbf{q}$ one obtains
\begin{align}
\begin{split}
	\hat{A}_\textbf{q} &= \hat{U}^\dagger \hat{a}_\textbf{q} \hat{U} = \int \sum_{n=0}^\infty \left[\mathrm{I}^{(2n)}_{\textbf{q},\textbf{q}'}\hat{a}_{\textbf{q}'} + \mathrm{I}^{(2n+1)}_{\textbf{q},\textbf{q}'}\hat{b}^\dagger_{\textbf{q}'}\right]\text{d}\textbf{q}',\\
	\hat{B}_\textbf{q} &= \hat{U}^\dagger \hat{b}_\textbf{q} \hat{U} = \int\sum_{n=0}^\infty \left[\mathrm{I}^{(2n)}_{\textbf{q},\textbf{q}'}\hat{b}_{\textbf{q}'} + \mathrm{I}^{(2n+1)}_{\textbf{q},\textbf{q}'}\hat{a}^\dagger_{\textbf{q}'}\right]\text{d}\textbf{q}',
\end{split}
\label{transform A y B}
\end{align}
in which the first terms of the expansion are
\begin{align}
\begin{split}
	\mathrm{I}^{(0)}_{\textbf{q}_1,\textbf{q}_2} &= \delta^{(2)}(\textbf{q}_1-\textbf{q}_2),\\
	\mathrm{I}^{(1)}_{\textbf{q}_1,\textbf{q}_2} &= \Gamma t \mathrm{F}(\textbf{q}_1,\textbf{q}_2),\\
	\mathrm{I}^{(2)}_{\textbf{q}_1,\textbf{q}_2} &= \dfrac{(\Gamma t)^2}{2!} \int \mathrm{F}(\textbf{q}_1,\textbf{q}')\mathrm{F}^*(\textbf{q}',\textbf{q}_2)\text{d}\textbf{q}',\\
	\mathrm{I}^{(3)}_{\textbf{q}_1,\textbf{q}_2} &= \dfrac{(\Gamma t)^3}{3!}\int \mathrm{F}(\textbf{q}_1,\textbf{q}')\mathrm{F}^*(\textbf{q}',\textbf{q}'')\mathrm{F}(\textbf{q}'',\textbf{q}_2)\text{d}\textbf{q}'\text{d}\textbf{q}'',
\end{split}
\label{coefficients}
\end{align}
The general structure of this series is as follows: the zeroth-order coefficient describes the transmission of the input beam in the corresponding mode, the first-order term represents the conversion of two pump photons into a probe--conjugate pair, and higher-order coefficients account for nested conversion processes mediated by the same TPA. Eqs.~\eqref{transform A y B} generalizes the input--output formalism of parametric amplification to an arbitrary spatial profile of the pump beam.

Two regimes can be distinguished from the structure of Eqs.~\eqref{transform A y B}. For a small coupling parameter $\Gamma$, the leading nontrivial coefficient $\mathrm{I}^{(1)}$ dominates, and the photon statistics are well approximated by truncating the series at the first order. This is the regime in which biphoton-state treatments of structured-pump parametric processes operate~\cite{Walborn2010, Boucher2021, Thachil2025, daMotta2025}, and it qualitatively captures the spatial transfer of the seed and pump structures to the generated fields. The full series is required whenever $\Gamma$ is sizable, and for any observable whose value depends quantitatively on the interplay of all orders, such as the exact gain, intensity-difference squeezing factor, magnitude of the spatial cross-correlation, and its position relative to the Cauchy--Schwarz bound. The closed-form summation enabled by the Dirac-comb expansion in Sec.~\ref{sec Dirac Comb} treats both regimes equally.

From the transformations in Eqs.~\eqref{transform A y B}, the photon-number operators per transverse wavevector $\hat{N}^{pr}_\textbf{q} = \hat{A}^\dagger_\textbf{q}\hat{A}_\textbf{q}$ and $\hat{N}^{c}_\textbf{q} = \hat{B}^\dagger_\textbf{q}\hat{B}_\textbf{q}$ can be evaluated for any input state. In particular, we consider stimulated 4WM, with the probe mode seeded with a coherent state while the conjugate mode is initialized in vacuum ($|\alpha\rangle_{pr}\otimes|0\rangle_c$). In the bright-beam limit, the vacuum contributions are negligible relative to the seeded contributions; thus, all expectation values below are understood in the normally ordered sense without writing the colons explicitly. In this regime the mean photon-number density per transverse wavevector is
\begin{align}
	\braket{ \hat{N}^{pr}_\textbf{q} } &= \int \sum_{n,m} \mathrm{I}^{(2n)*}_{\textbf{q},\textbf{q}'}\mathrm{I}^{(2m)}_{\textbf{q},\textbf{q}''}\braket{ \hat{a}^\dagger_{\textbf{q}'}\hat{a}_{\textbf{q}''} }_0\text{d}\textbf{q}'\text{d}\textbf{q}'',\\
	\braket{ \hat{N}^c_\textbf{q} } &= \int \sum_{n,m}\mathrm{I}^{(2n+1)*}_{\textbf{q},\textbf{q}'}\mathrm{I}^{(2m+1)}_{\textbf{q},\textbf{q}''}\braket{ \hat{a}^\dagger_{\textbf{q}''} \hat{a}_{\textbf{q}'} }_0 \text{d}\textbf{q}'\text{d}\textbf{q}'',
\end{align}
where $\left<\cdot\right>_0$ denotes the expectation value with respect to the input state of the process. These expressions yield the mean intensity in the far field for both beams.
\begin{widetext}
\noindent
Beyond the mean intensities, the same formalism yields the variances of each beam and the cross-covariance between them, where bipartite quantum correlations are encoded. Therefore, the variance of each field is
\begin{align}
    \braket{\Delta^2 \hat{N}^{pr}_{\textbf{q}_1,\textbf{q}_2}} &=\braket{\hat{N}^{pr}_{\textbf{q}_1}\hat{N}^{pr}_{\textbf{q}_2}} - \braket{\hat{N}^{pr}_{\textbf{q}_1}} \braket{\hat{N}^{pr}_{\textbf{q}_2}}\nonumber \\
    &= \int \sum_{nmlp} \left[\mathrm{I}^{(2n)*}_{\textbf{q}_1,\textbf{q}'}\mathrm{I}^{(2m)}_{\textbf{q}_2,\textbf{q}''}\mathrm{I}^{(2l)*}_{\textbf{q}_2,\textbf{q}'''}\mathrm{I}^{(2p)}_{\textbf{q}_1,\textbf{q}'''} + \mathrm{I}^{(2n)*}_{\textbf{q}_2,\textbf{q}'}\mathrm{I}^{(2m)}_{\textbf{q}_1,\textbf{q}''}\mathrm{I}^{(2l+1)*}_{\textbf{q}_1,\textbf{q}'''}\mathrm{I}^{(2p+1)}_{\textbf{q}_2,\textbf{q}'''}\right] \braket{\hat{a}^\dagger_{\textbf{q}'}\hat{a}_{\textbf{q}''}}_0\text{d}\textbf{q}'\text{d}\textbf{q}''\text{d}\textbf{q}''',\\
	\braket{ \Delta^2 \hat{N}^c_{\textbf{q}_1,\textbf{q}_2} } &=\braket{\hat{N}^{c}_{\textbf{q}_1}\hat{N}^{c}_{\textbf{q}_2}} - \braket{\hat{N}^{c}_{\textbf{q}_1}} \braket{\hat{N}^{c}_{\textbf{q}_2}}\nonumber \\
    &=\int \sum_{nmlp}\left[\mathrm{I}^{(2n+1)*}_{\textbf{q}_1,\textbf{q}'}\mathrm{I}^{(2m+1)}_{\textbf{q}_2,\textbf{q}''}\mathrm{I}^{(2l)*}_{\textbf{q}_2,\textbf{q}'''}\mathrm{I}^{(2p)}_{\textbf{q}_1,\textbf{q}'''} + \mathrm{I}^{(2n+1)*}_{\textbf{q}_2,\textbf{q}'}\mathrm{I}^{(2m+1)}_{\textbf{q}_1,\textbf{q}''}\mathrm{I}^{(2l+1)*}_{\textbf{q}_1,\textbf{q}'''}\mathrm{I}^{(2p+1)}_{\textbf{q}_2,\textbf{q}'''}\right] \langle\hat{a}^\dagger_{\textbf{q}'}\hat{a}_{\textbf{q}''}\rangle_0\text{d}\textbf{q}'\text{d}\textbf{q}''\text{d}\textbf{q}'''.
\end{align}

\begin{figure*}[!ht]
    \centering
    \includegraphics[width=5.648in]{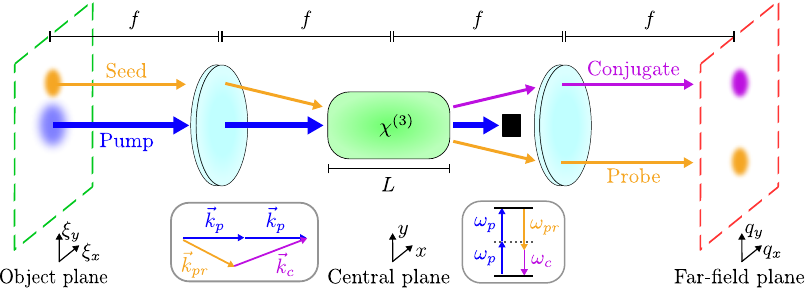}
    \caption{Setup for the propagation of arbitrary images in 4WM. An input image is imprinted on the pump and/or seed beams in the object plane defined by the axes ($\xi_x$, $\xi_y$). A Fourier-imaging lens maps the image to the center of the $\chi^{(3)}$ medium, where the 4WM interaction takes place. The second lens propagates the generated probe and conjugate beams to a far-field detection plane defined by the axes ($q_x$, $q_y$), where their intensities and cross-covariances are measured. The boxed insets illustrate energy conservation and the phase-matching condition.}
    \label{fig setup}
\end{figure*}
The covariance, which measures the spatial correlations of both beams, takes the form
\begin{align}
    \text{Cov}(\hat{N}^{pr}_{\textbf{q}_1},\hat{N}^{c}_{\textbf{q}_2})&= \braket{\hat{N}^{pr}_{\textbf{q}_1}\hat{N}^{c}_{\textbf{q}_2}} - \braket{\hat{N}^{pr}_{\textbf{q}_1}} \braket{\hat{N}^{c}_{\textbf{q}_2}}\nonumber \\
    &=\int \sum_{nmlp}\left[\mathrm{I}^{(2n)*}_{\textbf{q}_1,\textbf{q}'}\mathrm{I}^{(2m+1)*}_{\textbf{q}_2,\textbf{q}''}\mathrm{I}^{(2l)}_{\textbf{q}_1,\textbf{q}'''}\mathrm{I}^{(2p+1)}_{\textbf{q}_2,\textbf{q}'''}+ \mathrm{I}^{(2n+1)}_{\textbf{q}_2,\textbf{q}'}\mathrm{I}^{(2m)}_{\textbf{q}_1,\textbf{q}''}\mathrm{I}^{(2l)*}_{\textbf{q}_2,\textbf{q}'''}\mathrm{I}^{(2p+1)*}_{\textbf{q}_1,\textbf{q}'''}\right]\braket{\hat{a}^\dagger_{\textbf{q}'}\hat{a}_{\textbf{q}''}}_0\text{d}\textbf{q}'\text{d}\textbf{q}''\text{d}\textbf{q}''',
\end{align}

\end{widetext}
The observables derived above live in transverse-momentum (far-field) space, and are accessed by Fourier-imaging detection. The setup is shown in Fig.~\ref{fig setup}: the spatial profiles of the pump and seed are prepared in the object plane and Fourier transformed by a lens onto the center of the $\chi^{(3)}$ medium; after the interaction, a second lens maps the generated probe and conjugate to a far-field detection plane, where their intensities and cross-covariance are recorded.

\subsection{Intensity difference squeezing for multimode fields}

A standard figure of merit for bipartite quantum correlations is the intensity-difference squeezing (IDS), defined as the ratio of the noise of the intensity difference of the generated state to the shot-noise level of the equivalent uncorrelated two-mode coherent state~\cite{McCormick2006, Jasperse2011}:
\begin{equation}
	\text{IDS} = \dfrac{\langle\Delta^2 \hat{N}_-\rangle_{sq}}{\langle\Delta^2 \hat{N}_-\rangle_{coh}},
\end{equation}
with
\begin{align}
	\langle\Delta^2 \hat{N}_-\rangle_{sq} &= \iint\limits_{\textbf{q}_{pr},\textbf{q}_{pr}}\langle\Delta^2 \hat{N}^{pr}_{\textbf{q},\textbf{q}'}\rangle\text{d}\textbf{q}\text{d}\textbf{q}' + \iint\limits_{\textbf{q}_{c},\textbf{q}_{c}}\langle\Delta^2 \hat{N}^c_{\textbf{q},\textbf{q}'}\rangle\text{d}\textbf{q}\text{d}\textbf{q}'\nonumber \\
    &\hspace{3ex}- 2\iint\limits_{\textbf{q}_{c},\textbf{q}_{pr}} \text{Cov}(\hat{N}^{pr}_{\textbf{q}},\hat{N}^{c}_{\textbf{q}'})\text{d}\textbf{q}\text{d}\textbf{q}',
\end{align}
and
\begin{equation}
    \langle\Delta^2 \hat{N}_-\rangle_{coh} = \int\limits_{\textbf{q}_{pr}}\langle\hat{N}^{pr}_\textbf{q}\rangle\text{d}\textbf{q} + \int\limits_{\textbf{q}_{c}}\langle\hat{N}^{c}_\textbf{q}\rangle\text{d}\textbf{q},
\end{equation}
where the shot-noise denominator uses the fact that the two coherent reference modes are uncorrelated. In the variance terms, both integration variables run over the same detector region for each beam (the probe area for the probe variance and conjugate area for the conjugate variance). The above expressions evaluate the IDS over the entire transverse beam profile, as in typical bucket detector experiments. The same formalism returns a pixel-resolved IDS by restricting the integration domains in $\langle\Delta^2\hat{N}_-\rangle_{sq}$ and $\langle\Delta^2\hat{N}_-\rangle_{coh}$ to a single detector element, providing a position-dependent characterization of the correlations.

\subsection{Plane-Wave Pump}
As a consistency check of the formalism, we evaluate it in the limit of a plane-wave pump and recover the standard phase-insensitive amplifier results. The TPA function reduces to
\begin{equation}
	\mathrm{F}(\textbf{q}_1,\textbf{q}_2) = E_0^2\,\delta(\textbf{q}_1 + \textbf{q}_2).
\end{equation}
Assuming perfect phase-matching, the coefficients of the series take the form
\begin{equation}
	\mathrm{I}^{(n)}_{\textbf{q}_1,\textbf{q}_2} = \dfrac{s^{n}}{n!} \delta(\textbf{q}_1-(-1)^n\textbf{q}_2),
\end{equation}
where $s = (2\pi E_0)^2\Gamma t$ is the \textit{interaction strength}. The even and odd series of the coefficients are
\begin{align}
\begin{split}
	\sum_{n=0}^\infty \mathrm{I}^{(2n)}_{\textbf{q}_1,\textbf{q}_2} &= \cosh(s)\delta(\textbf{q}_1-\textbf{q}_2),\\
	\sum_{n=0}^\infty \mathrm{I}^{(2n+1)}_{\textbf{q}_1,\textbf{q}_2} &= \sinh(s)\delta(\textbf{q}_1+\textbf{q}_2).
\end{split}
\end{align}
Thus, the operators transform as follows
\begin{align}
\begin{split}
    \hat{A}_\textbf{q} &= \cosh(s)\hat{a}_\textbf{q} + \sinh(s)\hat{b}^\dagger_{-\textbf{q}},\\
    \hat{B}_\textbf{q} &= \cosh(s)\hat{b}_\textbf{q} + \sinh(s)\hat{a}^\dagger_{-\textbf{q}}.
\end{split}
\end{align}
For an arbitrary spatial distribution of the seed beam $\alpha_\textbf{q}$, normalized to $n_0 = \int |\alpha_\textbf{q}|^2\text{d}\textbf{q}$ input photons, the intensity distributions are given by
\begin{align}
    \langle\hat{N}^{pr}_\textbf{q}\rangle &=  G\left|\alpha_\textbf{q}\right|^2,\\
    \langle\hat{N}^{c}_\textbf{q}\rangle &=  (G-1)\left|\alpha_{-\textbf{q}}\right|^2,
\end{align}
with $G = \cosh^2(s)$ being the gain of the process. In this limit, the seed spatial profile is transferred to both the probe and conjugate intensities up to the gain factors $G$ and $G-1$, respectively. Integrated over the full transverse profile of each beam, the variances are expressed as
\begin{align}
    \langle\Delta^2 \hat{N}^{pr}\rangle &= (2G-1) G n_0,\\
    \langle\Delta^2 \hat{N}^{c}\rangle &= (2G-1) (G-1)n_0 ,
\end{align}
and the covariance is
\begin{equation}
    \text{Cov}(\hat{N}^{pr},\hat{N}^{c}) = 2G(G-1)n_0 .
\end{equation}
Therefore, the IDS is given by
\begin{equation}
    \text{IDS} = \dfrac{1}{2G-1}.
\end{equation}
This relates the IDS inversely to the amplification gain of the 4WM process.
The results recover the standard IDS for phase-insensitive parametric amplification in 4WM~\cite{Jasperse2011, Fang2015} and confirm the internal consistency of the formalism in the plane-wave limit.
\section{\label{sec Dirac Comb}Image propagation in 4WM}
\begin{figure}[t!]
    \centering
    \includegraphics[width=2.025in]{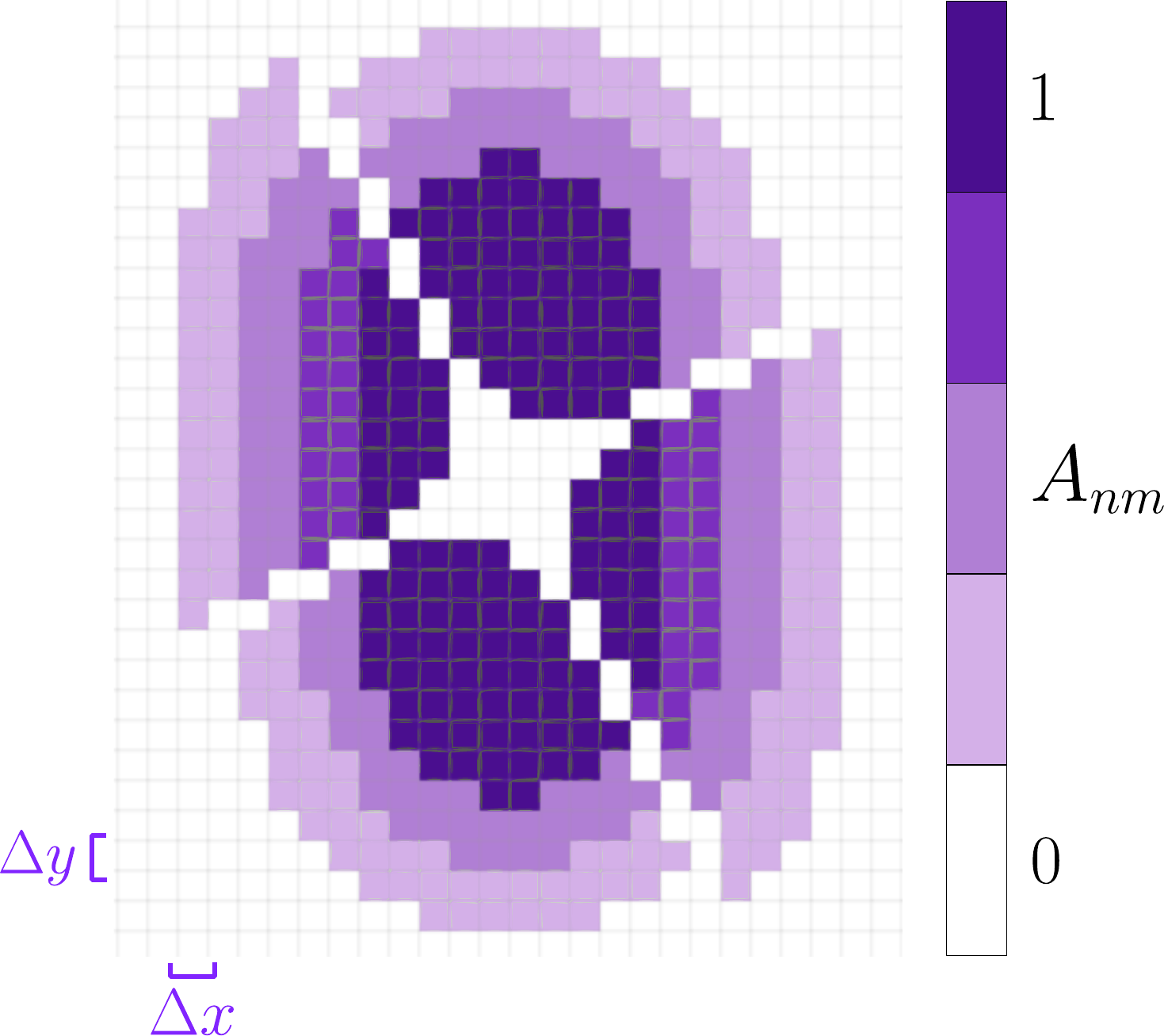}
    \caption{Illustration of the field's Dirac-comb expansion at the center of the medium, using the Amazing Quantum logo as an example.}
    \label{amazing field}
\end{figure}

We now apply the general framework to a pump field with an arbitrary transverse structure by expanding it in a Dirac comb as
\begin{equation}
	E(\boldsymbol{\rho}) = E_0\Delta x \Delta y \sum_{nm} A_{nm} \delta(\boldsymbol{\rho} - \boldsymbol{\rho}_{nm}),
\end{equation}
\noindent with $\boldsymbol{\rho}_{nm} = (x_n,y_m)$ the pixel positions and $A_{nm} = |A_{nm}|e^{i\phi_{nm}}$ the complex amplitude per pixel. The expression represents a grid of point sources with amplitudes and phases set by the two-dimensional matrix $A$, sampled on $N\times M$ pixels centered at $(x_{N/2},y_{M/2}) = (0,0)$. An example of this construction is illustrated in Fig.~\ref{amazing field}. We adopt the peak normalization $\max_{nm}|A_{nm}| = 1$, so that the dimensionless interaction strength $s = (2\pi E_0)^2\Gamma t$ corresponds to the local strength of the nonlinear interaction at the brightest pump pixel.
In this case, the TPA function is expressed as
\begin{equation}
	\mathrm{F}(\textbf{q}_1,\textbf{q}_2) = E_0^2 \frac{\Delta x \Delta y}{(2\pi)^2}\sum_{nm}A_{nm}^2 e^{-i (\textbf{q}_1 + \textbf{q}_2)\cdot \boldsymbol{\rho}_{nm}}.
\end{equation}

The coefficients of the expansion are
\begin{equation}
    \mathrm{I}^{(n)}_{\textbf{q}_1,\textbf{q}_2} = \frac{s^{n}}{n!}\frac{\Delta x\Delta y}{(2\pi)^2}\sum_{lp} |A_{lp}|^{2n} e^{-i (\textbf{q}_1 - (-1)^n \textbf{q}_2)\cdot \boldsymbol{\rho}_{lp}} e^{i\gamma_n \phi_{lp}},
\end{equation}
with $\gamma_{2n} = 0$ and $\gamma_{2n+1} = 2$. Then, the even and odd series of the coefficients are given by
\begin{align}
\begin{split}
	\sum_{n=0}^\infty \mathrm{I}^{(2n)}_{\textbf{q}_1,\textbf{q}_2} &= \frac{\Delta x\Delta y}{(2\pi)^2} \sum_{lp} \cosh(s |A_{lp}|^2) e^{-i (\textbf{q}_1 - \textbf{q}_2)\cdot \boldsymbol{\rho}_{lp}} ,\\
    \sum_{n=0}^\infty \mathrm{I}^{(2n+1)}_{\textbf{q}_1,\textbf{q}_2} &= \frac{\Delta x\Delta y}{(2\pi)^2} \sum_{lp} \sinh(s |A_{lp}|^2) e^{-i (\textbf{q}_1 + \textbf{q}_2)\cdot \boldsymbol{\rho}_{lp}} e^{i2\phi_{lp}}.
\end{split}
\end{align}
From the sum of these elements, the output transverse operators can be expressed as
\begin{align}
	\begin{split}
		\hat{A}_\textbf{q} &= \frac{\Delta x \Delta y}{2\pi} \sum_{nm} \hat{a}_{nm}\cosh\left( s |A_{nm}|^2 \right) e^{-i\textbf{q}\cdot \boldsymbol{\rho}_{nm}}\\
		&\hspace{3ex} + \frac{\Delta x \Delta y}{2\pi} \sum_{nm} \hat{b}^\dagger_{nm}\sinh\left( s |A_{nm}|^2 \right) e^{-i\textbf{q}\cdot \boldsymbol{\rho}_{nm}} e^{i2\phi_{nm}},\\
		\hat{B}_\textbf{q} &= \frac{\Delta x \Delta y}{2\pi} \sum_{nm} \hat{b}_{nm}\cosh\left( s |A_{nm}|^2 \right) e^{-i\textbf{q}\cdot \boldsymbol{\rho}_{nm}}\\
		&\hspace{3ex} + \frac{\Delta x \Delta y}{2\pi} \sum_{nm} \hat{a}^\dagger_{nm}\sinh\left( s |A_{nm}|^2 \right) e^{-i\textbf{q}\cdot \boldsymbol{\rho}_{nm}} e^{i2\phi_{nm}},
	\end{split}
\end{align}
where
\begin{align}
\begin{split}
	\hat{a}_{nm} &= \frac{1}{2\pi}\int \hat{a}_\textbf{q} e^{i\textbf{q}\cdot \boldsymbol{\rho}_{nm}} \text{d}^2q,\\
	\hat{b}_{nm} &= \frac{1}{2\pi}\int \hat{b}_\textbf{q} e^{i\textbf{q}\cdot \boldsymbol{\rho}_{nm}} \text{d}^2q,
\end{split}
\end{align}
are the spatial-representation annihilation operators at the pixel position $\boldsymbol{\rho}_{nm}$, associated with the field generated by the corresponding pump element. They satisfy the continuous-mode commutation relations $[\hat{a}_{nm}, \hat{a}^\dagger_{n'm'}] = \delta^{(2)}(\boldsymbol{\rho}_{nm}-\boldsymbol{\rho}_{n'm'})$ inherited from the field operators $\hat{a}_\textbf{q}$. Physically meaningful pixel observables involve the pixel-volume-weighted combinations $\Delta x\Delta y\,\hat{a}_{nm}$ that appear in the expressions below (see Appendix~\ref{app transformations dirac comb} for the derivation). The mean value of the output amplitudes is
\begin{align}
	\begin{split}
		\langle \hat{A}_\textbf{q}\rangle  &= \frac{\Delta x \Delta y}{2\pi} \sum_{nm} \alpha_{nm} \cosh\left( s |A_{nm}|^2 \right) e^{-i\textbf{q}\cdot \boldsymbol{\rho}_{nm}},\\
		\langle \hat{B}_\textbf{q}\rangle  &= \frac{\Delta x \Delta y}{2\pi} \sum_{nm} \alpha_{nm}^* \sinh\left( s |A_{nm}|^2 \right) e^{-i\textbf{q}\cdot \boldsymbol{\rho}_{nm}} e^{i2\phi_{nm}},
	\end{split}
\end{align}
where $\alpha_{nm} = \braket{\hat{a}_{nm}}_0$ is the seed in its spatial representation at the center of the medium. The seed amplitude is transferred to the probe and conjugate fields, which are weighted by hyperbolic functions of the pump amplitude. Thus, the mean intensities are
\begin{align}
	\braket{\hat{N}^{pr}_\textbf{q}} &= \left|\frac{\Delta x \Delta y}{2\pi}\sum_{nm} \alpha_{nm}\cosh\left( s |A_{nm}|^2 \right) e^{-i\textbf{q}\cdot \boldsymbol{\rho}_{nm}} \right|^2,\label{N probe}\\
	\braket{\hat{N}^c_\textbf{q}} &= \left| \frac{\Delta x \Delta y}{2\pi} \sum_{nm} \alpha_{nm}^* \sinh\left( s |A_{nm}|^2 \right) e^{-i\textbf{q}\cdot \boldsymbol{\rho}_{nm}}e^{i2\phi_{nm}}\right|^2,\label{N conjugate}
\end{align}
where the intensity of each field is the modulus squared of the angular spectrum of the seed weighted by the hyperbolic cosine or sine of the local pump amplitude. Hence, the mean value of the intensities describes the classical variable of the amplified fields. The quantum variables can then be computed by calculating the variance of this probe field. The probe variance is
\begin{align}
\begin{split}
    \braket{\Delta^2 \hat{N}^{pr}_{\textbf{q}_1,\textbf{q}_2}} &= \frac{(\Delta x \Delta y)^3}{(2\pi)^4}\sum_{nm} \alpha_{nm}^* \cosh\left( s |A_{nm}|^2 \right) e^{i \textbf{q}_1\cdot \boldsymbol{\rho}_{nm} } \\
	&\hspace{5ex}\times \sum_{nm} \alpha_{nm} \cosh\left( s |A_{nm}|^2 \right) e^{-i{\textbf{q}_2}\cdot \boldsymbol{\rho}_{nm}}\\
	&\hspace{5ex} \times \sum_{nm} \cosh^2\left( s |A_{nm}|^2 \right) e^{-i(\textbf{q}_1 - {\textbf{q}_2})\cdot \boldsymbol{\rho}_{nm}}\\
	&\hspace{3ex}+\text{c.c.},
\end{split}
\end{align}
and for the variance of the conjugate field
\begin{align}
\begin{split}
    \braket{\Delta^2 \hat{N}^{c}_{\textbf{q}_1,\textbf{q}_2}} &= \frac{(\Delta x \Delta y)^3}{(2\pi)^4}\sum_{nm} \alpha_{nm}^* \sinh\left( s |A_{nm}|^2 \right) e^{-i \textbf{q}_1\cdot \boldsymbol{\rho}_{nm} }e^{i2\phi_{nm}} \\
	&\hspace{5ex}\times \sum_{nm} \alpha_{nm} \sinh\left( s |A_{nm}|^2 \right) e^{i{\textbf{q}_2}\cdot \boldsymbol{\rho}_{nm}}e^{-i2\phi_{nm}}\\
	&\hspace{5ex} \times \sum_{nm} \sinh^2\left( s |A_{nm}|^2 \right) e^{i(\textbf{q}_1 - {\textbf{q}_2})\cdot \boldsymbol{\rho}_{nm}}\\
	&\hspace{3ex}+\text{c.c.}
\end{split}
\end{align}
Notably, the variances express the vacuum noise amplified by the seed and nonlinear interaction proportional to $\alpha_{nm}$. In addition, we determine the cross-covariance as follows
\begin{align}
\begin{split}
	\text{Cov}&(\hat{N}^{pr}_{\textbf{q}_1},\hat{N}^c_{{\textbf{q}_2}})\\
    &= \frac{(\Delta x \Delta y)^3}{(2\pi)^4}\sum_{nm} \alpha_{nm}^* \cosh\left( s |A_{nm}|^2 \right) e^{i \textbf{q}_1\cdot \boldsymbol{\rho}_{nm} } \\
	&\hspace{3ex}\times \sum_{nm} \alpha_{nm} \sinh\left( s |A_{nm}|^2 \right) e^{i{\textbf{q}_2}\cdot \boldsymbol{\rho}_{nm}}e^{-i2\phi_{nm}}\\
	&\hspace{3ex} \times \sum_{nm} \cosh\left( s |A_{nm}|^2 \right) \sinh\left( s |A_{nm}|^2 \right)\\
    &\hspace{13ex}\times e^{-i(\textbf{q}_1 + {\textbf{q}_2})\cdot \boldsymbol{\rho}_{nm}}e^{i2\phi_{nm}}\\
	&\hspace{3ex}+\text{c.c.}.
\end{split}
\end{align}
This is factorized into the product of the angular spectra of the probe and conjugate and the angular spectrum of a pump-dependent kernel. To define a figure of merit that measures the non-classicality of the correlations, we define the normalized correlation function as
\begin{equation}
    c_{\textbf{q}_1,\textbf{q}_2} = \dfrac{\text{Cov}(\hat{N}^{pr}_{\textbf{q}_1},\hat{N}^{c}_{\textbf{q}_2})}{\sqrt{\braket{\Delta^2 \hat{N}^{pr}_{\textbf{q}_1,\textbf{q}_1}} \braket{\Delta^2 \hat{N}^{c}_{\textbf{q}_2,\textbf{q}_2}}}},
\end{equation}
which is bounded by unity, according to the Cauchy-Schwarz inequality. Therefore, values larger than unity correspond to a violation of this inequality and are commonly interpreted as evidence of nonclassical correlations.

\subsection{Intensity imaging}

\begin{figure*}[t!]
    \centering
    \includegraphics[width=6.354in]{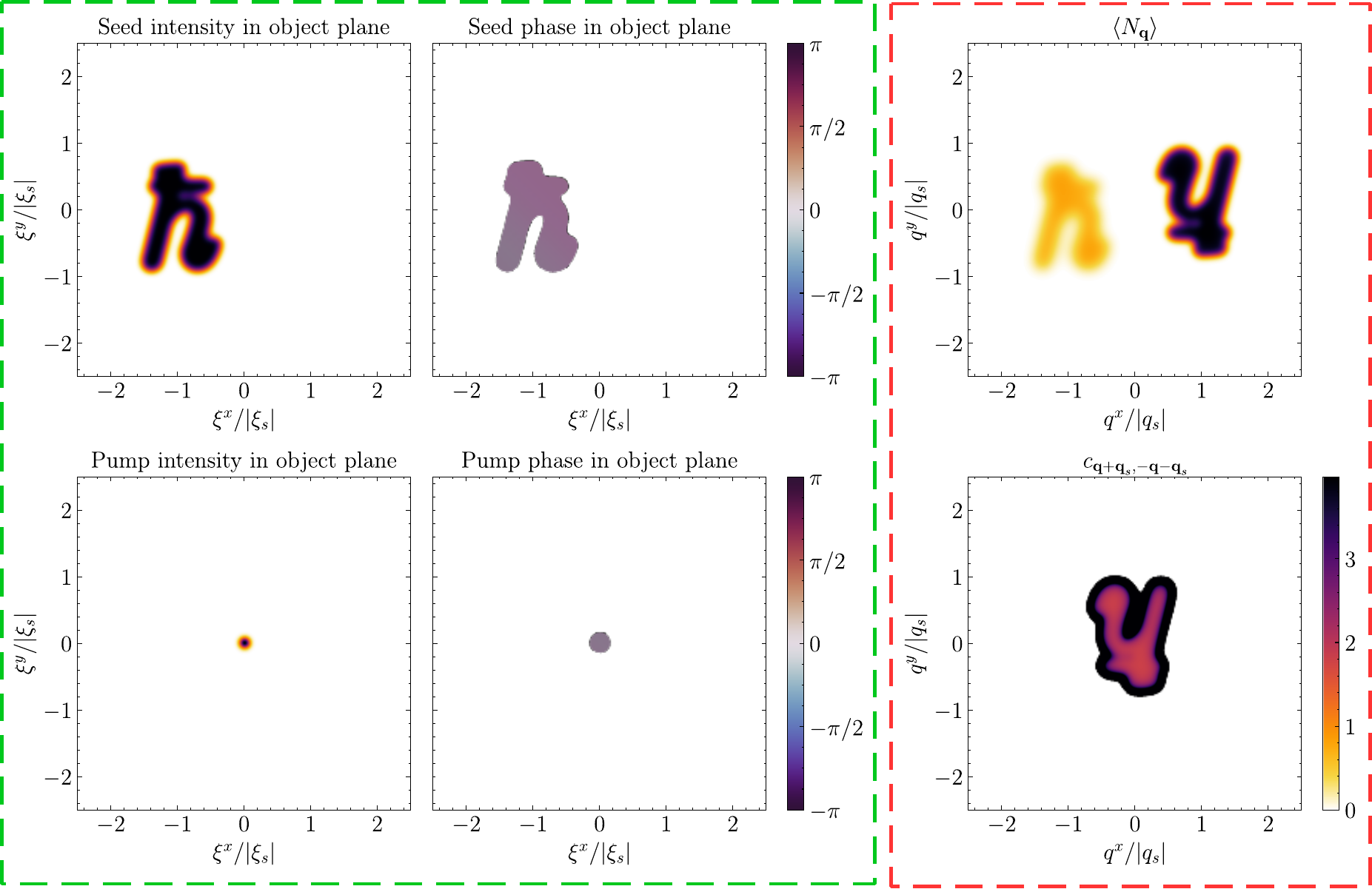}
    \caption{Intensity and phase profiles of the seed and pump fields in the $(\xi_x,\xi_y)$ object plane. The right column shows the intensity and normalized correlations in the $(q_x,q_y)$ far-field plane. The $\hbar$ symbol is imprinted in the object plane of the seed and mapped to its Fourier components at the center of the medium. The pump profile was Gaussian and spatially larger than the seed at the same plane, with a seed-to-pump width ratio $\delta = 0.38$. The $\hbar$ pattern appears in the far-field intensities $\braket{\hat{N}_\textbf{q}}$ of both the probe (right) and conjugate (left). The normalized correlation is evaluated at diametrically opposed momenta $\textbf{q}_{pr} = \textbf{q} + \textbf{q}_s = -\textbf{q}_c$ and also exhibits the $\hbar$ shape; all of its values exceed unity, violating the Cauchy--Schwarz inequality and witnessing nonclassical spatial correlations.}
    \label{fig hbar seed}
\end{figure*}

Earlier experiments have shown that, with a Gaussian pump, the bright probe and conjugate beams preserve the spatial structure of the seed~\cite{Boyer2008}. Our formalism reproduces and extends this picture. We consider an $\hbar$-shaped intensity profile for the seed and a Gaussian pump, both prepared in the object plane (Fig.~\ref{fig hbar seed}). After the lens, the seed propagates with a transverse wavevector $\textbf{q}_s$ and carries $n_0$ photons. Figure~\ref{fig hbar seed} shows the pump and seed field amplitudes and intensities in the $(\xi_x,\xi_y)$ object plane, as well as the intensity and normalized cross-correlation of the probe and conjugate in the $(q_x,q_y)$ far-field plane. The simulations were computed on a grid of $1001\times 1001$ pixels with a spacing $\Delta x = \Delta y = \pi/(3|\textbf{q}_s|)$.

The far-field intensities of the probe and conjugate preserve the $\hbar$-shaped structure of the seed. The conjugate appears as an inverted image of the probe, reflecting the phase-matching condition that correlates transverse wavevectors diametrically opposite with respect to the origin. We evaluate the covariance at opposite transverse momenta. In the low-interaction limit, for a Gaussian pump with constant phase ($\phi_{nm}=\,$const.), and $\delta\ll 1$, this reduces to
\begin{align}
\begin{split}
	\text{Cov}(\hat{N}^{pr}_{\textbf{q}_s+\textbf{q}},\hat{N}^c_{{-\textbf{q}_s-\textbf{q}}})&\approx 2s^2|A_{00}|^2\frac{(\Delta x \Delta y)^3}{(2\pi)^4} \sum_{nm} |A_{nm}|^2 \\
	&\hspace{3ex}\times \left|\sum_{nm} |\alpha_{nm}| e^{i\textbf{q}\cdot \boldsymbol{\rho}_{nm}}\right|^2,\nonumber 
\end{split}\\
    &\propto \left|\sum_{nm} |\alpha_{nm}| e^{i\textbf{q}\cdot \boldsymbol{\rho}_{nm}}\right|^2,
\end{align}
so that the covariance reduces to the modulus squared of the Fourier transform of the seed amplitude at the center of the medium (see Appendix~\ref{app covariance calculation} for derivation).
\begin{figure*}[t!]
    \centering
    \includegraphics[width=6.354in]{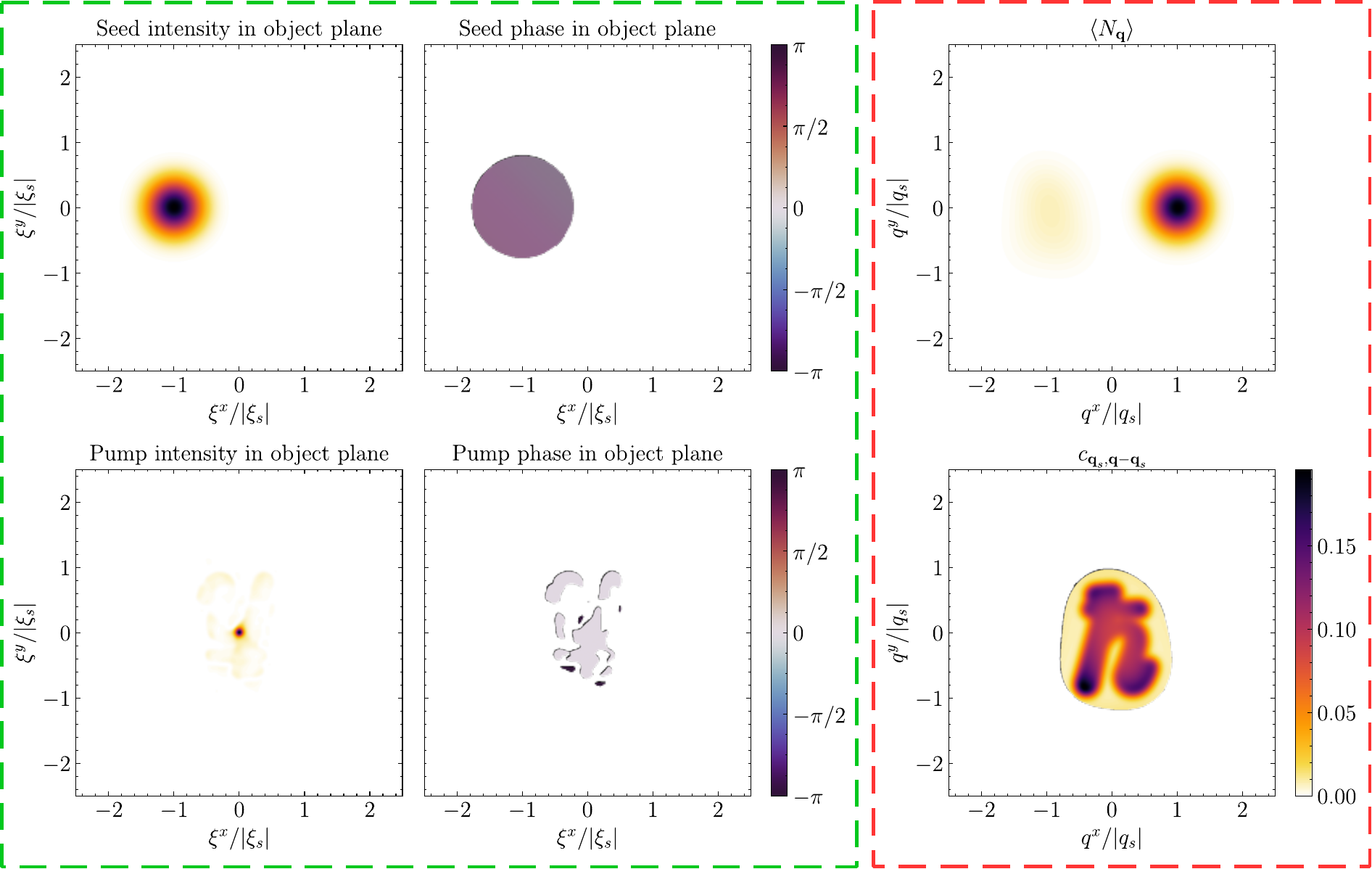}
    \caption{Covariance imaging. In the object plane, the seed is spatially Gaussian, and the pump has the shape required to reproduce the $\hbar$ symbol in the cross-correlation in the low-gain regime, with seed-to-pump width ratio $\delta = 0.21$. The probe and conjugate intensities (right and left Gaussian profiles, respectively) preserve the Gaussian structure of the seed. The normalized correlation, evaluated at $\textbf{q}_{pr} = \textbf{q}_s$ and $\textbf{q}_c = \textbf{q} - \textbf{q}_s$, reproduces the imprinted $\hbar$ pattern. All values of the normalized correlation remain below unity, so the Cauchy--Schwarz inequality is not violated by this configuration; this does not by itself establish classicality, which would require additional witnesses.}
    \label{fig cov hbar}
\end{figure*}

\subsection{Covariance imaging}
We now reverse the roles and imprint the $\hbar$ symbol in the spatial cross-correlation, a configuration we refer to as \textit{covariance imaging}. We take a Gaussian seed at the center of the medium,
with $\textbf{q}_s$ as the transverse wavevector. We evaluate the covariance at the center of the Gaussian probe profile while scanning over the conjugate momentum.
When the seed is spatially much smaller than the pump at the center of the cell, its angular spectrum is correspondingly broader~\cite{Kumar2021}, and in this limit, only the pixel $(n,m) = (0,0)$ contributes effectively to the sum involving $\alpha_{nm}$. The covariance then simplifies to
\begin{align}
\begin{split}
    \text{Cov}&(\hat{N}^{pr}_{\textbf{q}_s},\hat{N}^c_{{\textbf{q}-\textbf{q}_s}})\\
    &\propto \sum_{nm} \cosh\left( s |A_{nm}|^2 \right) \sinh\left( s |A_{nm}|^2 \right) e^{-i\textbf{q} \cdot \boldsymbol{\rho}_{nm}}e^{i2\phi_{nm}}\nonumber \\
    &\hspace{3ex}+\text{c.c.}
\end{split}
\end{align}
In the low-interaction limit $s\ll 1$ we obtain
\begin{align}
	\text{Cov}(\hat{N}^{pr}_{\textbf{q}_s},\hat{N}^c_{{\textbf{q}-\textbf{q}_s}}) \propto  \Re\left( \mathcal{F}[E^2](\textbf{q}) \right),
	\label{c(q)}
\end{align}
where we assume $A_{00}\neq 0$. This follows from taking $\delta\ll 1$. More precisely, the condition should be read as the central region of the pump on the scale of the seed support being non-vanishing. In this limit, the angular spectrum of the squared pump is transferred to the cross-correlation function~\cite{Nirala2023}.

Figure~\ref{fig cov hbar} shows the information transfer from the pump to the cross-correlation between probe and conjugate. A target $\hbar$ pattern is imprinted in the cross-correlation by taking the pump as
\begin{equation}
    E(\boldsymbol{\rho}) = \sqrt{\mathcal{F}^{-1}[\text{Cov}](\boldsymbol{\rho})}.
    \label{shape pump}
\end{equation}
By setting the target covariance to a real-valued $\hbar$-shaped profile, the pump field at the center of the medium is obtained from Eq.~\eqref{shape pump}. Figure~\ref{fig cov hbar} shows the seed (Gaussian) and pump intensity profiles in the object plane and at the center of the medium. The far-field probe and conjugate intensities remain approximately Gaussian, whereas their cross-covariance exhibits the imprinted $\hbar$ structure.


\subsection{Validity regime of the covariance-imaging recipe}
The pump-shaping prescription in Eq.~\eqref{shape pump} was derived in the limit $\delta\to 0$, $s\to 0$, where the angular spectrum of the squared pump is directly transferred to the cross-correlation. Outside this limit, the imprinted image is degraded by higher-order pump contributions to the photon statistics and by the finite size of the seed relative to the pump. We quantify the degradation by introducing a pixel-wise image-similarity score
\begin{equation}
	\mathcal{S} = 1 - \frac{1}{N M}\sum_{nm} \left| \frac{|c_\textbf{q}|}{\max |c_\textbf{q}|} - \frac{\Re\left( \mathcal{F}[E^2](\textbf{q}) \right)}{\max \Re\left( \mathcal{F}[E^2](\textbf{q}) \right)} \right|,
\end{equation}
defined as the normalized $L^{1}$ distance between the computed normalized correlation $|c_\textbf{q}|$ and the target pattern $\Re\{\mathcal{F}[E^2](\textbf{q})\}$. $\mathcal{S}=1$ corresponds to the perfect reproduction of the target, with decreasing values for stronger deviations. We adopt the term \textit{image similarity} rather than \textit{fidelity} to avoid conflict with the standard quantum-information meaning of the latter; structural similarity indices or other image-quality measures could be used equivalently and lead to the same qualitative conclusions.
\begin{figure}[t!]
    \centering
    \includegraphics[width=3.0375in]{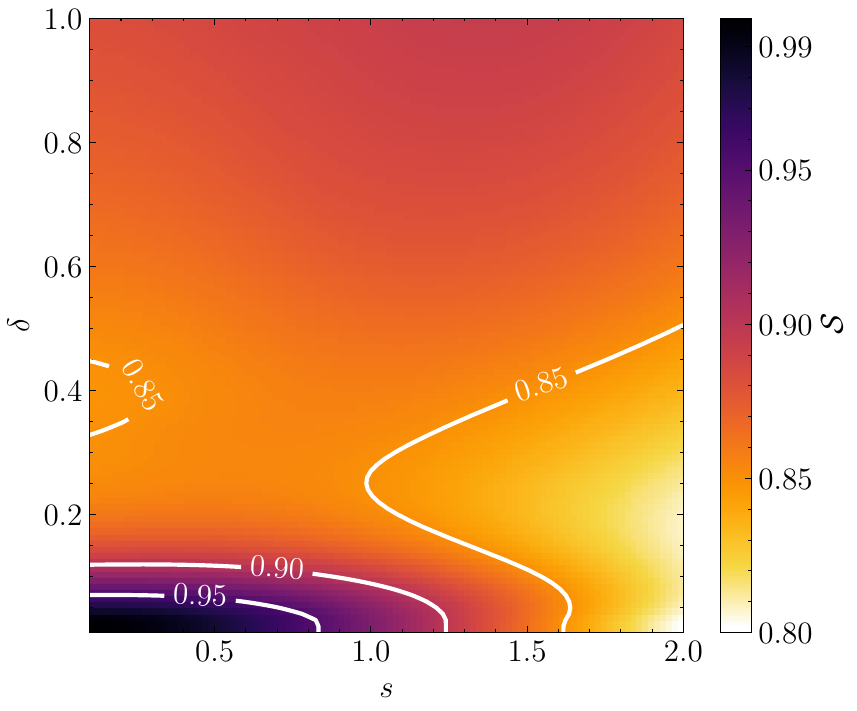}
    \caption{Image similarity $\mathcal{S}$ between the computed cross-correlation and the target $\hbar$ pattern, as a function of the seed-to-pump width ratio $\delta$ and the interaction strength $s$.}
    \label{fig fidelity}
\end{figure}

We define the transverse width of the pump and seed at the center of the medium as the standard deviation of the corresponding intensity profile, and $\delta$ as the ratio between the seed and pump widths. We evaluate $\mathcal{S}$ for an $\hbar$ target imprinted in the cross-correlation with a Gaussian seed by scanning $\delta$ and $s$.

The results are shown in Fig. ~\ref{fig fidelity}. The high-similarity region is concentrated in the lower-left corner of the $(\delta, s)$ plane. For $\delta \lesssim 0.16$ and $s \lesssim 1.2$ we find $\mathcal{S} \gtrsim 0.9$, and the imprinted $\hbar$ pattern is faithfully recovered in the cross-correlation. The similarity decreases monotonically with $s$, reflecting the growing weight of higher-order pump contributions (involving four or more photons) that deviate from the first-order relation between the pump profile and the cross-correlation used to derive Eq.~\eqref{shape pump}. It also decreases with $\delta$, because the seed then samples a non-uniform $|A_{nm}|^2$ across its support, and the substitution $A_{nm}\to A_{00}$ used in the derivation breaks down.

The dependence of $\mathcal{S}$ on $\delta$ is non-monotonic: the similarity decreases up to $\delta \approx 0.3$--$0.4$, where the assumption $A_{nm}\to A_{00}$ is maximally violated, and then increases again as $\delta$ grows toward unity and beyond. In the second regime, the seed becomes spatially wide enough to act effectively as a plane wave over the support of the pump, restoring a high-fidelity transfer of the squared-pump angular spectrum to the cross-correlation. 

For larger $(\delta, s)$ values, the closed-form inversion of Eq.~\eqref{shape pump} is no longer accurate, as the relationship between the pump profile and imprinted covariance becomes intrinsically nonlinear in both the parameters. A numerical inversion of the full all-order expressions in Sec.~\ref{sec model} would yield, in principle, the pump profile to produce a target covariance for arbitrary $(\delta, s)$; this extension is left for future studies.
\section{\label{sec conclusions}Conclusions}
We developed a macroscopic continuous-variable model of stimulated four-wave mixing that yields closed-form expressions for the mean intensity, variance, and cross-covariance of the bright probe and conjugate beams, valid for arbitrary transverse profiles of the pump and seed and to all orders in the dimensionless interaction strength. The model is built on a Dirac-comb expansion of the pump field, which renders the photon statistics computable, pixel by pixel, from a single set of expressions. In the plane-wave limit, it reduces to the standard phase-insensitive amplifier results, providing a consistency check for the formalism.

Beyond this limit, the same formalism makes explicit the spatial routing of information observed in earlier bright-twin-beam experiments. With a Gaussian pump and an arbitrarily shaped seed, the seed image is transferred to the far-field intensities of the probe and conjugate, in agreement with the seed-imaging experiments of Ref.~\cite{Boyer2008}. With a Gaussian seed and a structured pump, the angular spectrum of the squared pump field is transferred to the spatial cross-correlation between the two beams, providing a macroscopic all-order theoretical account of the covariance-imaging experiments of Ref.~\cite{Nirala2023}.

For covariance imaging, the pump-shaping prescription in Eq.~\eqref{shape pump} is derived from a low-order analytical inversion and is therefore restricted to the parameter window characterized by the image similarity score introduced in Sec.~\ref{sec Dirac Comb}. The recipe reproduces a target pattern with high similarity when $\delta \ll 1$ and $s \lesssim 1$, where the pump is locally close to a plane wave, and higher-order pump contributions remain subdominant. Outside this regime, the relationship between the pump profile and the imprinted covariance becomes intrinsically nonlinear in both parameters; a numerical inversion of the all-order expressions in Sec.~\ref{sec model} yields the pump to carve any target covariance for arbitrary $(\delta, s)$.

Beyond the configurations analyzed here, the same formalism applies to arbitrary combinations of structured pump and structured seed, to input states beyond coherent seeds, and to 4WM in cold atomic ensembles, where the all-order regime is experimentally accessible~\cite{Lambrecht1996, Arajo2022}.

{\it Acknowledgments.---} The authors thank Dr.~Timothy S. Woodworth for valuable input and discussions during the early stages of this project. This work was partially supported by the CIFAR Catalyst Fund CF-0435-CP24-069; FONDECYT Grants No. 1230897, No. 1240204; and ANID - Millennium Science Initiative Program Grant No. ICN17-012. H.M.F. acknowledges the São Paulo Research Foundation (FAPESP) Grant No. 2024/22385-6, and 2024/08522-0 for financial support.

\bibliography{FWM_bib}

\newpage
\appendix
\begin{widetext}
    \section{\label{app transformations dirac comb} Detailed calculation of the expansion for a Dirac comb pump}
We explicitly compute the coefficients of the series. At order zero,
\begin{equation}
	\mathrm{I}^{(0)}_{\textbf{q}_1,\textbf{q}_2} = \delta(\textbf{q}_1 - \textbf{q}_2).
\end{equation}
At first order,
\begin{align}
	\mathrm{I}^{(1)}_{\textbf{q}_1,\textbf{q}_2} &= \Gamma t E_0^2 \frac{\Delta x^2 \Delta y^2}{(2\pi)^2}\sum_{nmlp}A_{nm}A_{lp}e^{-i (\textbf{q}_1 + \textbf{q}_2)\cdot \boldsymbol{\rho}_{lp}}\int e^{-i \textbf{q}'\cdot (\boldsymbol{\rho}_{nm}-\boldsymbol{\rho}_{lp})} \text{d}^2 q'\nonumber\\
	&= \Gamma t E_0^2 \frac{\Delta x^2 \Delta y^2}{(2\pi)^2}\sum_{nmlp}A_{nm}A_{lp}e^{-i (\textbf{q}_1 + \textbf{q}_2) \cdot \boldsymbol{\rho}_{lp}} \frac{(2\pi)^2}{\Delta x\Delta y} \delta_{nl}\delta_{mp}\nonumber\\
	&= s \frac{\Delta x \Delta y}{(2\pi)^2}\sum_{nm}A_{nm}^2 e^{-i (\textbf{q}_1 + \textbf{q}_2) \cdot \boldsymbol{\rho}_{nm}} ,
\end{align}
with $s = (2\pi E_0)^2\Gamma t$ being the interaction strength. At second order,
\begin{align}
	\mathrm{I}^{(2)}_{\textbf{q}_1,\textbf{q}_2} &= \frac{s^2}{2} \frac{\Delta x^2\Delta y^2}{(2\pi)^4}\sum_{nmlp}A_{nm}^2A_{lp}^{*2} e^{-i \textbf{q}_1 \cdot \boldsymbol{\rho}_{nm}} e^{i \textbf{q}_2 \cdot \boldsymbol{\rho}_{lp}}\int e^{-i \textbf{q}'\cdot (\boldsymbol{\rho}_{nm}-\boldsymbol{\rho}_{lp})} \text{d}^2 q' \nonumber\\
	&= \frac{s^2}{2} \frac{\Delta x\Delta y}{(2\pi)^2}\sum_{nm} |A_{nm}|^{4} e^{-i (\textbf{q}_1 - \textbf{q}_2) \cdot \boldsymbol{\rho}_{nm}}.
\end{align}
At third order,
\begin{align}
	\mathrm{I}^{(3)}_{\textbf{q}_1,\textbf{q}_2} &= \frac{s^3}{3!} \frac{\Delta x^2\Delta y^2}{(2\pi)^4}\sum_{nmlp} |A_{nm}|^{4}A_{lp}^2 e^{-i \textbf{q}_1 \cdot \boldsymbol{\rho}_{nm}} e^{-i \textbf{q}_2 \cdot \boldsymbol{\rho}_{lp}} \int e^{i \textbf{q}'\cdot (\boldsymbol{\rho}_{nm}-\boldsymbol{\rho}_{lp})} \text{d}^2 q'\nonumber\\
	&= \frac{s^3}{3!} \frac{\Delta x\Delta y}{(2\pi)^2}\sum_{nm} |A_{nm}|^{4}A_{nm}^2 e^{-i (\textbf{q}_1 + \textbf{q}_2) \cdot \boldsymbol{\rho}_{nm}}.
\end{align}
The terms computed thus far are sufficient to identify the coefficient patterns. Splitting the series into even and odd terms yields
\begin{align}
	\mathrm{I}^{(2n)}_{\textbf{q}_1,\textbf{q}_2} &= \frac{s^{{2n}}}{(2n)!}\frac{\Delta x\Delta y}{(2\pi)^2}\sum_{lp} |A_{lp}|^{4n} e^{-i (\textbf{q}_1 - \textbf{q}_2) \cdot \boldsymbol{\rho}_{lp}}\\
	\mathrm{I}^{(2n+1)}_{\textbf{q}_1,\textbf{q}_2} &= \frac{s^{{2n+1}}}{(2n+1)!}\frac{\Delta x\Delta y}{(2\pi)^2}\sum_{lp} |A_{lp}|^{4n}A_{lp}^2 e^{-i (\textbf{q}_1 + \textbf{q}_2) \cdot \boldsymbol{\rho}_{lp}}
\end{align}
Summing over all $n$,
\begin{align}
	\sum_{n=0}^\infty \mathrm{I}^{(2n)}_{\textbf{q}_1,\textbf{q}_2} &= \delta(\textbf{q}_1 - \textbf{q}_2) + \frac{\Delta x\Delta y}{(2\pi)^2} \sum_{lp} \left[\cosh(s |A_{lp}|^2)-1\right] e^{-i (\textbf{q}_1 - \textbf{q}_2) \cdot \boldsymbol{\rho}_{lp}}\nonumber\\
	&\approx \frac{\Delta x\Delta y}{(2\pi)^2} \sum_{lp} \cosh(s |A_{lp}|^2) e^{-i (\textbf{q}_1 - \textbf{q}_2) \cdot \boldsymbol{\rho}_{lp}},
\end{align}
where we have used the approximation
\begin{equation}
	\frac{\Delta x\Delta y}{(2\pi)^2} \sum_{lp} e^{-i (\textbf{q}_1 - \textbf{q}_2) \cdot \boldsymbol{\rho}_{lp}} \approx \delta(\textbf{q}_1 - \textbf{q}_2),
\end{equation}
valid for a sufficiently fine grid size. Strictly, this sum produces a comb of shifted Dirac delta distributions; however, when restricted to the central region (the first Brillouin zone of the grid), the other contributions are absent. Proceeding analogously with the odd series,
\begin{equation}
	\sum_{n=0}^\infty \mathrm{I}^{(2n+1)}_{\textbf{q}_1,\textbf{q}_2} = \frac{\Delta x\Delta y}{(2\pi)^2} \sum_{lp} \sinh(s |A_{lp}|^2) e^{-i (\textbf{q}_1 + \textbf{q}_2) \cdot \boldsymbol{\rho}_{lp}}e^{i2\phi_{lp}},
\end{equation}
with $A_{lp} = |A_{lp}| e^{i\phi_{lp}}$. Defining the spatial-representation operators
\begin{align}
\begin{split}
	\hat{a}_{nm} &= \frac{1}{2\pi}\int \hat{a}_\textbf{q} e^{i\textbf{q}\cdot \boldsymbol{\rho}_{nm}} \text{d}^2q,\\
	\hat{b}_{nm} &= \frac{1}{2\pi}\int \hat{b}_\textbf{q} e^{i\textbf{q}\cdot \boldsymbol{\rho}_{nm}} \text{d}^2q,
\end{split}
\end{align}
the annihilation operators transform as
\begin{align}
	\begin{split}
		\hat{A}_\textbf{q} &= \frac{\Delta x \Delta y}{2\pi} \sum_{nm} \hat{a}_{nm}\cosh\left( s |A_{nm}|^2 \right) e^{-i\textbf{q}\cdot \boldsymbol{\rho}_{nm}}  + \frac{\Delta x \Delta y}{2\pi} \sum_{nm} \hat{b}^\dagger_{nm}\sinh\left( s |A_{nm}|^2 \right) e^{-i\textbf{q}\cdot \boldsymbol{\rho}_{nm}} e^{i2\phi_{nm}},\\
		\hat{B}_\textbf{q} &= \frac{\Delta x \Delta y}{2\pi} \sum_{nm} \hat{b}_{nm}\cosh\left( s |A_{nm}|^2 \right) e^{-i\textbf{q}\cdot \boldsymbol{\rho}_{nm}} + \frac{\Delta x \Delta y}{2\pi} \sum_{nm} \hat{a}^\dagger_{nm}\sinh\left( s |A_{nm}|^2 \right) e^{-i\textbf{q}\cdot \boldsymbol{\rho}_{nm}} e^{i2\phi_{nm}},
	\end{split}
\end{align}
whose mean values, for a seed prepared in a coherent state whose spatial distribution in the position representation is $\alpha(x_n,y_m) = \alpha_{nm}$, is
\begin{align}
	\begin{split}
		\langle \hat{A}_\textbf{q}\rangle  &= \frac{\Delta x \Delta y}{2\pi} \sum_{nm} \alpha_{nm} \cosh\left( s |A_{nm}|^2 \right) e^{-i\textbf{q}\cdot \boldsymbol{\rho}_{nm}},\\
		\langle \hat{B}_\textbf{q}\rangle  &= \frac{\Delta x \Delta y}{2\pi} \sum_{nm} \alpha_{nm}^* \sinh\left( s |A_{nm}|^2 \right) e^{-i\textbf{q}\cdot \boldsymbol{\rho}_{nm}} e^{i2\phi_{nm}}.
	\end{split}
\end{align}

\end{widetext}

\begin{widetext}
    \section{\label{app covariance calculation} Detailed calculation of the covariances}

The conjugate appears as an inverted image of the probe, reflecting the phase-matching condition that correlates transverse wavevectors diametrically opposite with respect to the origin. We evaluate the covariance at opposite transverse momenta as follows:
\begin{align}
\begin{split}
	\text{Cov}&(\hat{N}^{pr}_{\textbf{q}_s+\textbf{q}},\hat{N}^c_{{-\textbf{q}_s-\textbf{q}}})= \frac{(\Delta x \Delta y)^3}{(2\pi)^4}\sum_{nm} |\alpha_{nm}| \cosh\left( s |A_{nm}|^2 \right) e^{i \textbf{q}\cdot \boldsymbol{\rho}_{nm} } \\
	&\hspace{3ex}\times \sum_{nm} |\alpha_{nm}| \sinh\left( s |A_{nm}|^2 \right) e^{-i\textbf{q}\cdot \boldsymbol{\rho}_{nm}}e^{-i2\phi_{nm}} \sum_{nm} \cosh\left( s |A_{nm}|^2 \right) \sinh\left( s |A_{nm}|^2 \right) e^{i2\phi_{nm}}+\text{c.c.},
\end{split}
\end{align}
In the low-interaction limit and for a Gaussian pump with a constant phase ($\phi_{nm}=\,$const.), this reduces to
\begin{align}
\begin{split}
	\text{Cov}(\hat{N}^{pr}_{\textbf{q}_s+\textbf{q}},\hat{N}^c_{{-\textbf{q}_s-\textbf{q}}})&\approx s^2\frac{(\Delta x \Delta y)^3}{(2\pi)^4}\sum_{nm} |\alpha_{nm}| e^{i \textbf{q}\cdot \boldsymbol{\rho}_{nm} }\sum_{nm} |\alpha_{nm}|  |A_{nm}|^2e^{-i\textbf{q}\cdot \boldsymbol{\rho}_{nm}}\sum_{nm} |A_{nm}|^2 +\text{c.c.},
\end{split}
\end{align}
and if $\delta\ll 1$ the variation of $|A_{nm}|^2$ over the seed support is negligible and we can substitute $A_{nm}\to A_{00}$, giving
\begin{align}
	\text{Cov}(\hat{N}^{pr}_{\textbf{q}_s+\textbf{q}},\hat{N}^c_{{-\textbf{q}_s-\textbf{q}}})&\approx 2s^2|A_{00}|^2\frac{(\Delta x \Delta y)^3}{(2\pi)^4} \sum_{nm} |A_{nm}|^2 \left|\sum_{nm} |\alpha_{nm}| e^{i\textbf{q}\cdot \boldsymbol{\rho}_{nm}}\right|^2 
\propto \left|\sum_{nm} |\alpha_{nm}| e^{i\textbf{q}\cdot \boldsymbol{\rho}_{nm}}\right|^2,
\end{align}
so that the covariance reduces to the modulus squared of the Fourier transform of the seed amplitude at the center of the medium.
\end{widetext}

\end{document}